\documentclass[aps,prb,twocolumn,nopacs,floats,superscriptaddress]{revtex4-2}
\usepackage{lineno}       
\usepackage{hyperref}     
\usepackage{amsmath,amssymb} 
\usepackage{graphicx}     
\usepackage{float}        

\begin{document}

\title{Bose one-component plasma in two dimensions: A Monte Carlo study}

\author{Massimo Boninsegni}
\affiliation{Department of Physics, University of Alberta, Edmonton, Alberta, Canada T6G 2H5}
\affiliation{Phenikaa Institute for Advanced Study, Phenikaa University, Nguyen Trac Street, Duong Noi Ward, Hanoi, Vietnam}

\begin{abstract}
The low-temperature properties of a 2D Bose fluid of charged particles interacting through a $1/r$ potential, moving in the presence of a uniform neutralizing background, is studied by Quantum Monte Carlo simulations. We make use of the Modified Periodic Coulomb potential formalism to 
account for the long-range character of the interaction, and explore a range of density corresponding to average interparticle separation $1\le r_s\le 80$. We report numerical results based on simulations of system comprising up to 2304 particles. We find a superfluid ground state for $r_s$ as large as 70, i.e., significantly above the most recent numerical estimate of the Wigner crystallization threshold
. Furthermore, no thermally re-entrant crystalline phase nor any evidence of metastable bubbles are observed near the transition, in contrast with a previous theoretical study in which quantum statistics was neglected. Altogether, the computed superfluid transition temperature depends remarkably weakly on density. 
\end{abstract}

\date{\today}
\maketitle
\section{Introduction}
A fluid of identical charged particles moving in the presence of a uniform, electrically neutralizing background of opposite charge, is of broad interest in different areas of condensed matter physics. This model is sometimes referred to as {\em Jellium}, or one-component plasma (OCP), the latter being the terminology that we shall adopt here. If particles obey Fermi statistics, the OCP represents the simplest model to capture the basic physics of electrons in metals, for example (see, for instance, Ref. \cite{fetter1971}). The case of Bose statistics, however, has significance as well, as it may be relevant to Fermi systems in which the interaction leads to the formation of relatively tightly bound pairs, each pair acting essentially as a boson. In two dimensions (2D), the Bose OCP is speculated to capture at least some of the physics of layered superconductors, especially in the context of bipolaron theories \cite{alexandrov1995,Marchand2010,Sous2018,Zhang2022}.

We consider here the 2D Bose OCP, modeled as an ensemble of $N$ identical, electrically charged particles of mass $m$ and spin zero, enclosed in a square box of side $L$, with periodic boundary conditions in both directions. The quantum-mechanical Hamiltonian, expressed in dimensionless units, reads as follows:
\begin{equation}\label{ham}
\hat h = -\frac{1}{2}\sum_i\nabla^2+\sum_{i<j}\frac{1}{|{\bf r}_i-{\bf r}_j|}
\end{equation}
where ${\bf r}_i$ is the position of the $i$th particle, all lengths are expressed in terms of the Bohr radius $a$ \cite{fetter1971}, and $\epsilon\equiv \hbar^2/(ma^2)$, usually referred to as the Hartree, is the unit of energy and temperature (i.e., we set the Boltzmann constant $k_B=1$). The nominal system density is $n\equiv N/L^2$, but it is customary to utilize the dimensionless mean interparticle distance $r_s$, defined through $n\equiv1/(\pi r_s^2 a^2)$. It is also convenient, for reasons that will become clear later, to define at the outset the (density-dependent) characteristic temperature $T^\star\equiv \epsilon/r_s^2$.

In the high density ($r_s\to 0)$ limit, the interparticle interactions have negligible effect, and the system approaches the non-interacting Bose gas limit. A number of analytical theories have been developed for this limit \cite{hines1979,sim1986,um1990}, in which the system displays a superfluid (superconducting) ground state. As the density is lowered (i.e., for $r_s > 1$) the potential energy of interaction increasingly dominates over the kinetic energy; in this strongly interacting regime, approximate analytical approaches are less dependable, making Quantum Monte Carlo (QMC) simulations a 
valid computational alternative, as the system features Bose statistics and therefore there is no ``sign'' problem. Indeed, the phase diagram of this system has been investigated by QMC both at zero \cite{depalo2004} and finite temperature \cite{clark2009}. 
For a sufficiently large value of $r_s$ the system is expected to undergo solidification at low temperature to a non-superfluid ground state featuring crystalline long-range order (also known as Wigner crystal). The most recent estimate of the value of $r_s$ at which this takes place (at temperature $T=0$) is $r_W\approx$ 66 \cite{clark2009}. 

In principle, no ordinary first-order phase transition can occur, even at temperature $T=0$, in a many-body system featuring pairwise interactions falling off with distance as $1/r$ \cite{spivak2004}. Specifically, the coexistence of two phases (crystal and superfluid) separated by a macroscopic interface is energetically unfavorable, as the energy can be lowered by the formation of a ``microemulsion", with large solid clusters (``bubbles'') floating in the superfluid. 
While the underlying mathematical argument is rigorous, the bubbles have a characteristic size $W_0$ which is related to physical quantities like the interfacial energy and the difference in chemical potential between the coexisting phases. In Ref. \cite{clark2009}, an order-of-magnitude estimate for $W_0$ was provided, showing that it is too large (much larger than essentially any imaginable physical system) for such a scenario to be relevant in practice (the same conclusion was reached in the context of a 2D purely repulsive dipolar gas \cite{moroni2014}, for which the same prediction had originally been made). However, metastable bubbles of a size of the order of a few times the interparticle distance were observed in the simulations of Ref. \cite{clark2009}, together with a finite-temperature re-entrant crystalline phase near the transition. 

An important point to note is that the calculation of Ref. \cite{clark2009} did not include effects of quantum statistics, i.e., particles were regarded as {\em distinguishable}. However, subsequent theoretical work has shown that quantum exchanges can significantly affect the location of the phase boundary, as exchanges impart thermodynamic stability to the liquid phase; for example, they are ultimately responsible for the failure of $^4$He to crystallize at low, but finite temperature \cite{boninsegni2012}. 

In order to assess the robustness of the existing theoretical predictions, in the present work we carried out a study of the same model using a finite-temperature QMC methodology allowing us to incorporate in full quantum statistics. We computed the superfluid response of the system as a function of temperature and $r_s$, using a cooling protocol allowing us to detect the crystallization of the system at intermediate temperature, i.e., the possible presence of a re-entrant crystalline phase. We also computed ground state energetic and structural properties and compared them here with the results of Ref. \cite{depalo2004}.

The main results of our investigation are the following:
\begin{enumerate}
    \item{The largest value of $r_s$ for which a superfluid transition is observed at low temperature is $r_s=70$, which is significantly above the estimated Wigner crystallization threshold ($r_W=66$) of Ref. \cite{clark2009}. We estimate $r_W\sim 71$. The determined superfluid transition temperature $T_c(r_s)$ is well approximated by $\sim 0.75\ T^\star$ in entire domain of existence of the superfluid phase, a fact already noticed in Ref. \cite{zhang2023}. On approaching the Wigner crystallization, $T_c$ plateaus at $\sim 0.6 \ T^\star$.}
    \item {The re-entrant crystalline phase reported in Ref. \cite{clark2009} is not observed in this work. Such a re-entrance is likely an artifact of the neglect of quantum exchanges, which are fully included in our calculation. We also see no evidence of the formation of crystalline bubbles in the fluid phase near crystallization, also reported in Ref. \cite{clark2009}. }
    \item{The comparison of computed ground state properties with those of Ref. \cite{depalo2004} shows reasonable quantitative agreement, despite the fact that different methodologies are utilized, including the way in which the long-range part of the Coulomb interaction is treated.}
\end{enumerate}
\section{Methods}\label{meth}
The calculations carried out in this work are all based on the canonical implementation \cite{mezzacapo2006,mezzacapo2006b} of the continuous-space Worm Algorithm \cite{boninsegni2006,boninsegni2006b}, a finite temperature QMC simulation method based on Feynman's space-time formulation of quantum statistical mechanics \cite{Feynman1965}.  This methodology is extensively described in the literature, hence it will not be reviewed here. 
At least for Bose systems, it delivers results that can be regarded as numerically ``exact'', i.e., statistical and systematic errors can be rendered negligible in practice, using ordinary computing infrastructure. 
As mentioned above, the effect of quantum statistics are entirely included, i.e., we do {\em not} treat particles as distinguishable. 

An important aspect, when one is attempting to simulate a system of particles interacting via a long-range potential, like in the case of this study, is the computational scheme through which the effect of the interaction beyond the finite simulation cell is incorporated. The traditional choice for 1/$r$ systems is the Ewald summation method (ESM) \cite{ewald1921,wells2015,wang2019}, which involves a significant computational overhead.  In this work, we made use of the considerably simpler, Modified Periodic Coulomb scheme (MPC) of Fraser {\it et al.} \cite{fraser1996}. While in principle both the ESM and the MPC schemes converge to the same result in the thermodynamic limit, the MPC affords a substantial computational speedup over the ESM. Although rather detailed comparisons of the two methods have been published \cite{williamson1997,drummond2008,kolorenc2011,azadi2015}, to our knowledge they have been mostly focused on electronic, not Bose systems. Thus, one of the goals of this research work is to assess the reliability of the MPC with charged Bose fluids.

Other details of the simulation are  standard; the short-time approximation to the imaginary-time propagator used here is the one commonly known as ``primitive'', which is accurate to second order in the time step $\tau$. We carried out numerical extrapolation of the estimates to the $\tau\to 0$ limit, and generally observed convergence of the thermal averages for a value of $\tau\approx {10^{-3}}\ {T^\star}^{-1}$ for all quantities of interest here. These include the energy per particle, the pair correlation function and the superfluid fraction $\rho_S(T)$, which we compute by means of the standard winding number estimator \cite{pollock1987}. 

Our simulated systems comprise a number $N$ of particles ranging from 36 to 2304. In order to carry out a temperature scan for fixed values of $N, r_s$, we typically start from a temperature high enough so that the system can be established to be in a fluid phase, and the proceed to cool down in steps, on each dividing the temperature by two. Each cooling step  requires to double the number of time slices (see, for instance, Ref. \cite{boninsegni2006b}); this can be accomplished fairly efficiently by first interpolating intermediate particle positions along the many-particle paths of a configuration generated at the higher temperature, and then allowing the system to equilibrate at the lower one.

\section{Results}
\subsection{Ground state}
As mentioned in Section \ref{meth}, the calculations carried out here make use of a finite temperature methodology, i.e., obtaining ground state estimates requires in principle to carry out an extrapolation to the $T=0$ limit of estimates obtained at finite temperature. In concrete terms, that means reaching a temperature sufficiently low (typically $\lesssim 0.3\  T^\star$) that the physical estimates no longer change appreciably, within the statistical errors of the calculation.

\begin{figure}[H]
    \centering
    \includegraphics[width=1.0\linewidth]{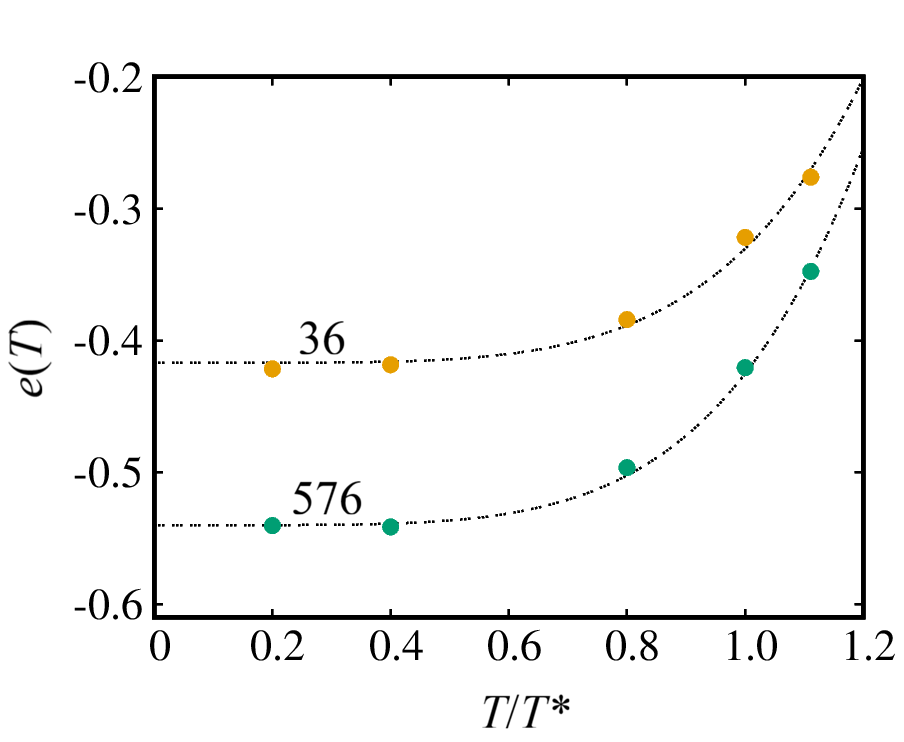}
    \caption{Computed energy per particle (in Hartree) as a function of temperature for a system with $r_s=1$. Data are shown for two different system sizes, comprising $N=36$ and $N=576$ particles. Lines through the data are fits obtained as explained in the text. Statistical errors are smaller than symbol sizes.}
    \label{eofT}
\end{figure}

An example of this procedure is shown in Fig. \ref{eofT} for the energy per particle $e(T)$ computed as a function of temperature (expressed in units of $T^\star$), for a system with $r_s=1$. The solid lines represent fits to the data based on the expression $e(T)=a+bT^5$, consistent with a 2D plasmon dispersion $\omega_k\propto k^{1/2}$. Henceforth, whenever characterizing results as ``ground state'' we shall imply that a procedure such as that shown in Fig. \ref {eofT} was carried out.
 
\begin{figure}[H]
    \centering
    \includegraphics[width=1.0\linewidth]{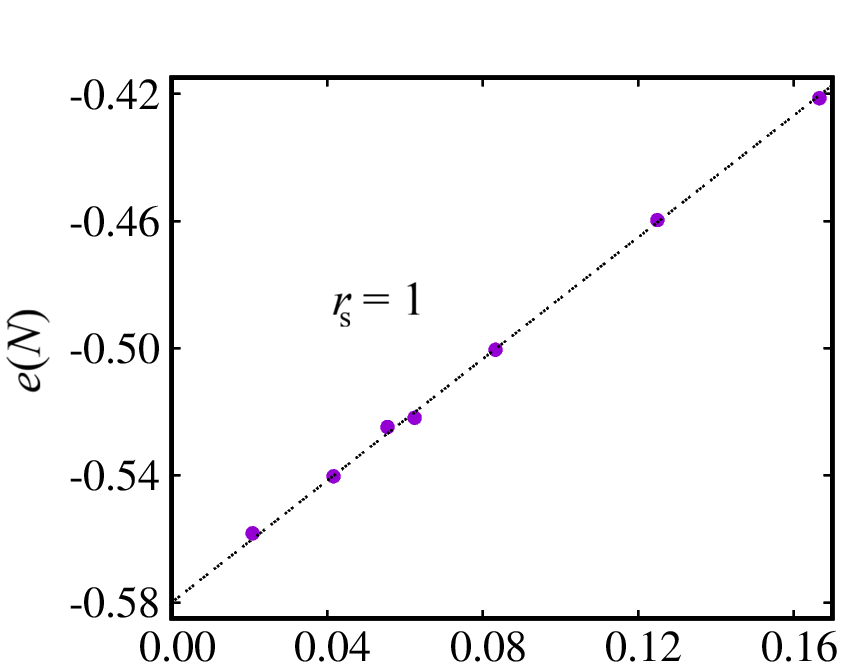}
        \includegraphics[width=1.0\linewidth]{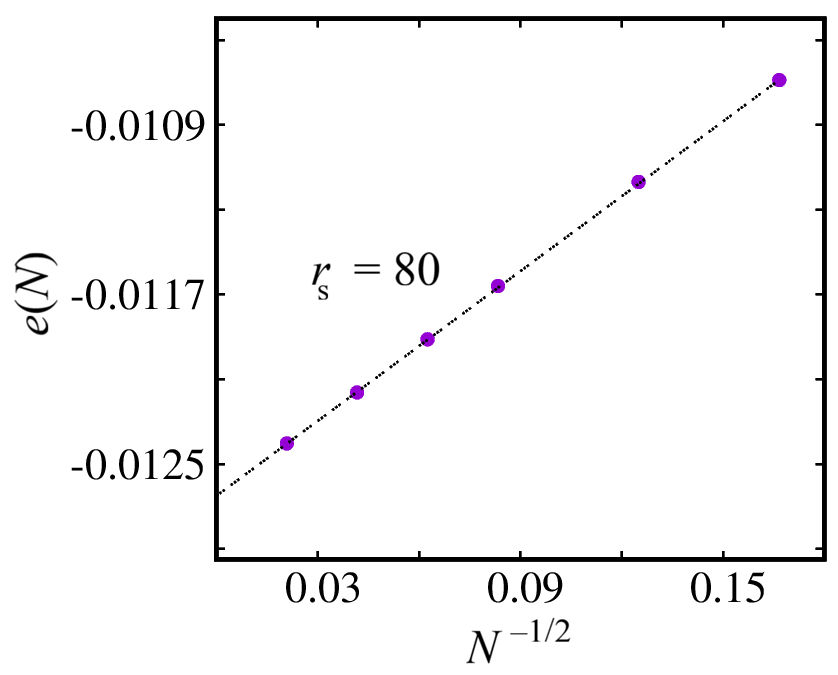}
    \caption{Extrapolation of the ground state energy per particle (in Hartree) to the thermodynamic limit for two different values of $r_s$, namely $r_s=1$ (top) and $r_s=80$ (bottom). Statistical errors are smaller than symbol sizes.}
    \label{extrapolation}
\end{figure}

Our aim is also to obtain results valid in the thermodynamic limit, i.e., $N,L\to \infty$ while keeping $n$ constant. Bogoliubov theory of the weakly interacting Bose gas, expected to be valid in the $r_s\to 0$ limit, points to a size dependence of the ground state energy per particle $e_0(N)=e_0(\infty)+\lambda N^{-1/2}$. We find this expression to provide an {\em excellent} fit for our energy data in the low-temperature limit for {\em all} values of $r_s$ considered in this work, including where the ground state is a crystal, i.e., for $r_s=80$. This is illustrated in Fig. \ref{extrapolation}. 

Our ground state estimates for the total ($e$) and kinetic ($k$) energy  per particle are reported in Table \ref {table} for several values of $r_s$. It should be mentioned that the size dependence of the ground state energy reflects that of the  potential energy; this is not surprising, as the kinetic energy is mainly affected by the local environment experienced by the particles. Indeed, the kinetic energy per particle is found to be size independent (within statistical uncertainties) for $N > 64$, essentially for {\em all} values of $r_s$. Thus, the results quoted in Table \ref{table} for $k$ are those obtained for a system of $N=2304$ particles.
\\ \indent 
We compare our results against the Diffusion Monte Carlo (DMC) results of Ref. \cite{depalo2004}. The quantitative agreement between the two calculations is quite satisfactory, despite the difference in simulated system sizes (an order of magnitude greater in the present work, which obviously benefits from over two decades of advances in computing infrastructure), and the fact that different methods were utilized to incorporate the effect of the long range of the Coulomb interaction (the ESM was used in Ref. \cite{depalo2004}). 
\\ \indent 
The DMC method, which is specifically designed for the ground state, typically delivers results with considerably smaller statistical uncertainties than finite temperature techniques, but is often prone to systematic errors arising from the trial wave function and the finite projection time and size of the population of walkers utilized \cite{moroni2012,boninsegni2001} (it should be mentioned that there exists other ground state QMC methods, e.g. the Reptation Quantum Monte Carlo \cite{baroni1999} and the Path Integral Ground State \cite{sarsa2000,cuervo2005}  that are free of this problem, as they are not based on a population of walkers). In light of that, the quantitative agreement of the kinetic energy estimates, which as mentioned above are not very much affected by the long-range part of the interaction, is certainly heartening.
\\ \indent
\begin{table}[h]
\centering
\begin{tabular} {|c|c|c|c|c| }\hline
$r_s$ & $e$ (This work) & $e$ (Ref. \cite{depalo2004}) &$k$ (This work) &$k$ (Ref. \cite{depalo2004}) \\ \hline
1 & $-0.5800(10)$  & $-0.5724(2)$ &$0.142(2)$ &$0.145$ \\ 
2 & $-0.3388(9)$  & $-0.3370(1)$ &$0.0717(8)$ & $0.0721$\\
5 & $-0.1597(1)$ & $-0.159151(25)$ &$0.0243(1)$ & $0.02448$\\
10 &$-0.08746(3)$ &$-0.08740(2)$ &$0.00974(5)$ & $0.009805$ \\
20 &$-0.04683(3)$ & $-0.046694(4)$ &$0.00376(4)$ & $0.00376$\\
48 &$-0.02062(2)$ & & & \\
64 &$-0.015688(9)$ & & & \\
72 &$ -0.014011(6)$ & & & \\
80 & $-0.012650(6)$ & & & \\
\hline
\end{tabular}
\caption{Ground state total ($e$) and kinetic ($k$) energy per particle (in Hartree) as a function of the parameter $r_s$, extrapolated to the thermodynamic limit. They are compared to the DMC results of Ref. \cite{depalo2004}. Statistical errors (in parentheses) are on the last digits. For the kinetic energy, the values quoted pertain to a system with $N=2304$ particles. No statistical errors are provided for $k$ in Ref. \cite{depalo2004}.}\label{table}
\end{table}

\begin{figure}[H]
    \centering
    \includegraphics[width=1.0\linewidth]{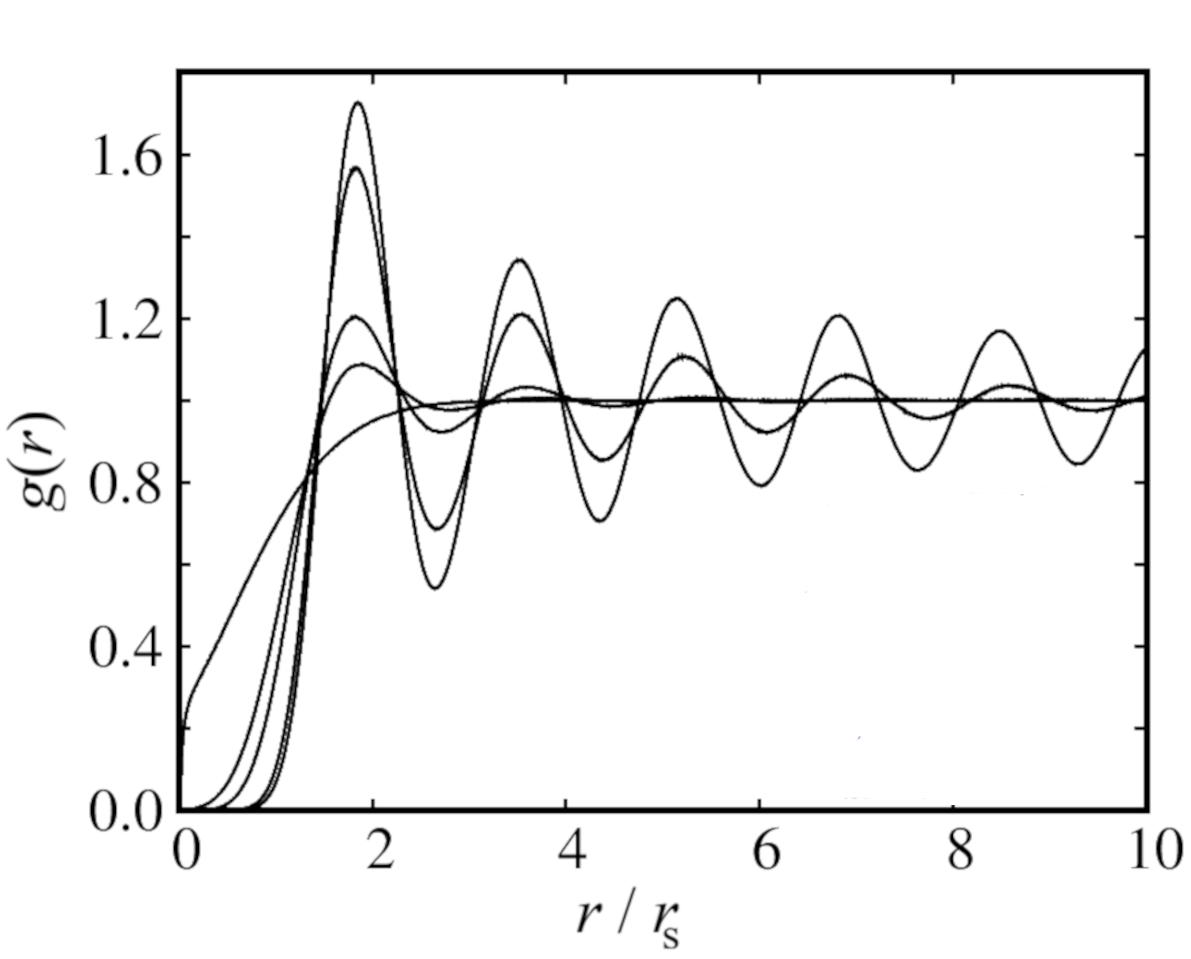}
    \caption{Pair correlation function computed in the low temperature limit for $r_s=1, 10, 20, 68, 80$. Higher values of $r_s$ correspond to higher peaks. 
    }
    \label{gofr}
\end{figure}

Fig. \ref{gofr} shows the pair correlation function computed at low temperature for various values of $r_s$. The behavior is qualitatively that which is expected, with increasingly persistent oscillations as the system approaches crystallization at low density. As discussed below, of all the curves shown in Fig. \ref{gofr} only that for $r_s=80$ (whose main peak exceeds 1.6 in height) corresponds to a crystalline ground state. A comparison with the results of Ref. \cite{depalo2004} shows broad agreement. 

The pair correlation function and the related static structure factor are the quantities of choice to establish the presence of crystalline long-range order, signaled by persistent oscillations in the $g(r)$, or, equivalently, a sharp peak in the static structure factor. As mentioned above, we find no evidence of crystalline order in the low temperature limit for $r_s$ up to 70, which is the largest value for which a superfluid transition is observed (we come back to this point below). This allows us to locate Wigner crystallization in the neighborhood of $r_s = r_W \approx 71$.
As mentioned above, in Ref. \cite{depalo2004}, $r_W$ was estimated at $\approx 60$; that estimate was subsequently revised upward [to $\approx 66$, specifically 66.5(2)] in Ref. \cite{clark2009}; the considerably greater system sizes utilized therein, allowing for a more reliable extrapolation to the thermodynamic limit, seem to provide a plausible explanation for the discrepancy between the two calculations. In Ref. \cite{clark2009}, a re-entrant crystalline phase was also observed at finite temperature, in correspondence of a non-crystalline (i.e., fluid) ground state. 

Our results differ in two ways from those of Ref. \cite{clark2009}, namely we find that the ground state is a (super)fluid for $r_s$ at least up to 70 and, more importantly, we see {\em no evidence} of any low-temperature thermo-crystallization, i.e., of re-entrant solid phases. We establish this conclusion both from the direct visual inspection of the many-particle configurations generated by our algorithm  for a given value for $r_s$, as well as by monitoring the pair correlation function and the superfluid fraction on varying the temperature. For example, for $r_s=68$ the $g(r)$ remains virtually unchanged as the temperature is lowered from $2\ T^\star$ to $0.5\ T^\star$, i.e., a temperature range in which the system transitions from a normal fluid to a superfluid (see below), without any concomitant structural change. 

The neglect of quantum-mechanical exchanges of identical particles in the study of \cite{clark2009} is at the root of the 
the discrepancy between our findings and theirs, near the liquid-solid boundary. While it is accepted that exchanges are generally suppressed in the crystal phase, they render the superfluid phase energetically  competitive with respect to the crystalline one. Conversely, their exclusion strengthens the crystalline phase, leading to unphysical thermo-crystallization at finite temperature. as shown in Ref. \cite{boninsegni2012} for $^4$He.\\
Likewise, we did not observe any consistent evidence of solid bubbles occurring in the liquid phase, reported in Ref. \cite{clark2009}, either by visual inspection of the many-particle configurations (a procedure often subjective and ambiguous in any case), nor through a careful examination of correlation functions, such as the $g(r)$.  Rather, the Bose liquid is observed to be disordered and homogeneous, and in this case too it seems that bubbles may arise merely as a result of the absence of quantum-mechanical exchanges.

\subsection{Superfluid  transition}

In the low $r_s$ limit, i.e., at high density, the ground state of the system is a homogeneous Bose superfluid, displaying a Berezinskii-Kosterlitz-Thouless transition \cite{berezinskii1971,berezinskii1972,kosterlitz1972,kosterlitz1973} at finite temperature, one of the best characterized theoretically. One of the most important results is the so-called {\em universal jump} condition \cite{nelson1977} relating the superfluid transition temperature $T_c$ to the value of the superfluid fraction at $T_c$, specifically \begin{equation}\rho_S(T_c)=(T_c/T^\star),\end{equation} in the units used here. The conventional wisdom is that as a system approaches crystallization its superfluid response is suppressed, as a result of stronger interparticle interactions. In bulk 3D $^4$He, for example, the superfluid transition temperature is reduced by approximately 20\% if the system is pressurized to approximately 25 bars, i.e.near freezing. However, a recent study of the Bose OCP \cite{zhang2023} has shown that the ratio $T_c/T^\star$ is a non-monotonic function of $r_s$, peaking at $r_s\sim 20$ and then relatively slowly decreasing as the density is lowered toward Wigner crystallization. In this work we explored the superfluid behavior of the system near the phase boundary.
\\ \indent
As mentioned above, we observe a superfluid transition for $r_s$ as high as 70, as we lower the temperature through the cooling protocol described above. In particular, we find $\rho_S = 0.68(4)$ at $T=0.624\ T^\star$. On the other hand, no evidence of superfluidity is observed for $r_s=72$, down to the lowest temperature reached in this study, namely $T=0.34\ T^\star$, for the system of largest size simulated in this work (i.e., $N=2304$); consistently, the only exchange cycles that are observed involve a handful of particles at the most.
\\ \indent 
\begin{figure}[H]
    \centering
    \includegraphics[width=0.9\linewidth]{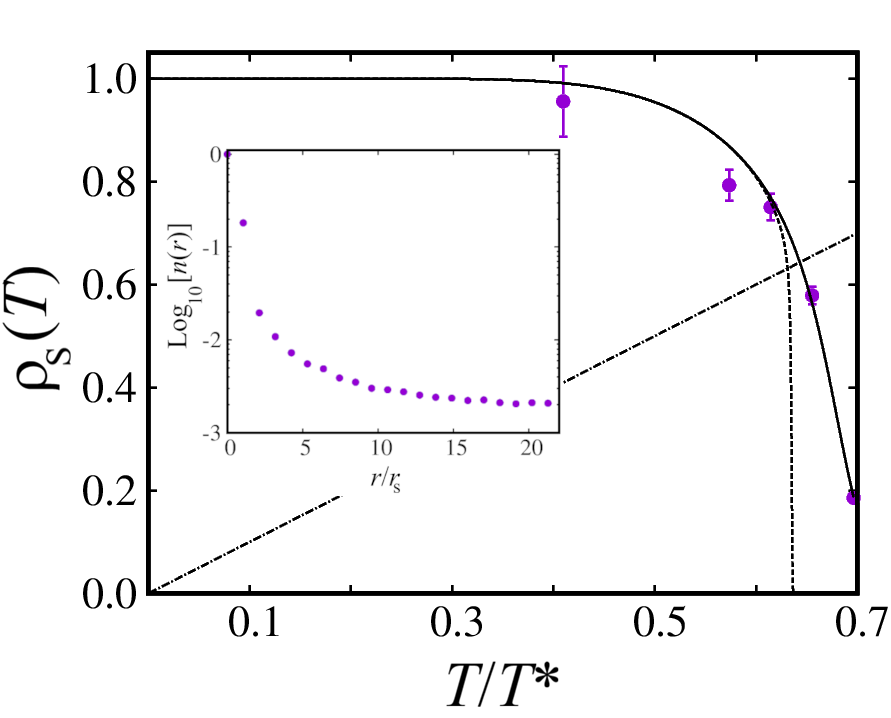}
    \caption{Superfluid fraction $\rho_S(T)$ for $r_s=64$ for a system comprising $N=576$ particles. When not shown, statistical errors are smaller than symbol size. Solid line through the data is obtained by solving the recursive KT equations (see text), whereas the dotted line represents the extrapolation to the thermodynamic limit. Dotted-dashed line corresponds to $T/T^\star=1$. Inset shows the one-body density matrix $n(r)$ computed at $T/T^\star=0.41$. }
    \label{rhost}
\end{figure}

In order to determine the superfluid transition temperature we follow the procedure outlined in Ref. \cite{PhysRevB.39.2084}, i.e., we fit the computed values of the superfluid fraction $\rho_S(T)$ by solving the recursive Kosterlitz-Thouless equations, using the two fitting parameters (vortex energy and diameter) to extrapolate the curve $\rho_S(T)$ to the thermodynamic limit \cite{boninsegni1999}. Fig. \ref{rhost} shows an example of this procedure for the case $r_s=64$. Also shown (inset) is the computed one-body density matrix at the lowest temperature, namely $T=0.41 \ T^\star$.
\begin{figure}[H]
    \centering
    \includegraphics[width=0.9\linewidth]{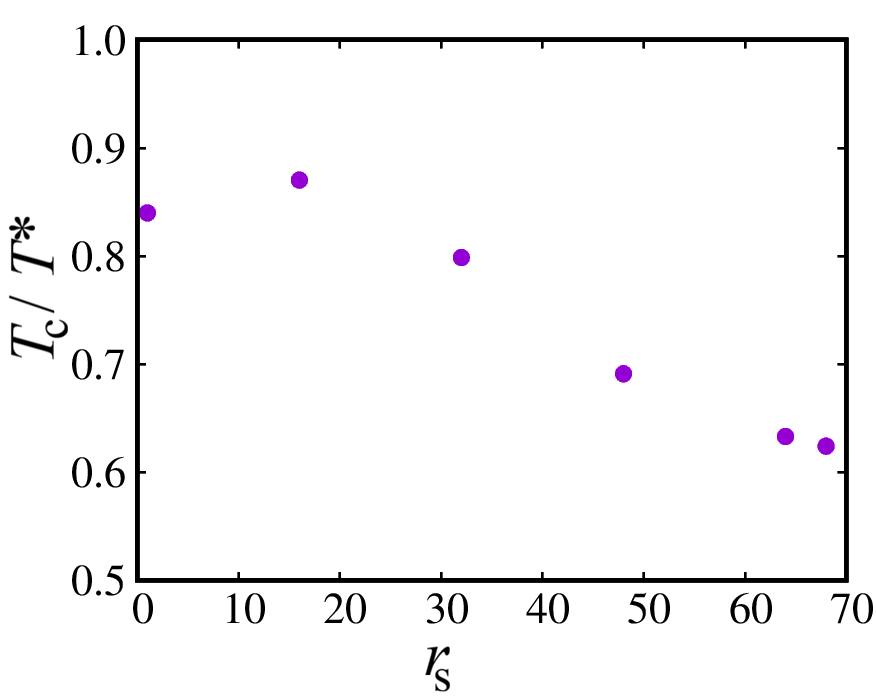}
    \caption{Superfluid transition temperature $T_c$ (in units of $T^\star$) computed as a function of $r_s$. }
    \label{tc}
\end{figure}
Our results for the superfluid transition temperature $T_c$ as a function of $r_s$ are illustrated in Fig. \ref{tc}. The most remarkable aspect, already pointed out in Ref. \cite{zhang2023} is the very mild dependency of the transition temperature on the density. Indeed, over the whole range of densities explored in this work, including the near vicinity of Wigner crystallization, $T_c$ falls within the relatively narrow range $0.6-0.9\ T^\star$, the maximum attained near $r_s=20$.

It is interesting to note that this is not quantitatively dissimilar from what is observed in the archetypal superfluid system, namely $^4$He, in 2D. In that case too, there is a comparable fractional increase ($\sim\ 50\ \%$) of the superfluid transition temperature and of the density on compressing the system from equilibrium (i.e., at saturated vapor pressure) all the way to freezing \cite{gordillo1998}.

\section{Conclusion}
We have carried out an extensive numerical investigation of the low temperature phase diagram of the Bose OCP in 2D, at low temperature. This model is believed to be relevant to different scenarios of high-temperature superconductivity, chiefly those based on (bi)polarons. We used a QMC methodology that differs from those utilized in previous studies, with the twofold aim of providing an independent check of previous results, as well as to furnish reliable numerical predictions for quantities for which they did not yet exist.

Our ground state results are largely consistent with those of the study by De Palo {\em et. al.} \cite{depalo2004}. In our view, the small quantitative differences can be ascribed to the different sizes of the simulated systems, as well as the way in which the long-range tail of the interaction is treated and, possibly, the used of ``mixed'' estimators in the calculation of ground state expectation values for quantities other thant he energy in Ref. \cite{depalo2004}. The most significant difference is the location of the Wigner crystallization, which we place at approximately $r_W\approx71$, as opposed to $r_s\approx 60$ in Ref. \cite{depalo2004}.

More significant are the differences with the finite-temperature calculation of Ref. \cite{clark2009}, which places Wigner crystallization at $r_s\approx 66$, i.e., closer to our estimate, but also reports metastable solid bubbles in the fluid phase, as well as a re-entrant crystalline phase at finite temperature, of which we find no evidence here. We believe both to be spurious effects of the neglect of quantum-mechanical exchanges in Ref. \cite{clark2009}. Altogether, the phase diagram of this system appears rather conventional, with no evidence of exotic phases (e.g., ``supersolid'', which are believed to require soft-core type interactions \cite{boninsegni2012b,kora2019}).

This work was supported by the Natural Sciences and Engineering Research Council of Canada, under grant RGPIN 2024-05664. The author wishes to acknowledge useful conversations with N. V. Prokof'ev. 
\bibliography{references}

\end{document}